\documentclass[prl,twocolumn,amsmath,amssymb,english,superscriptaddress,longbibliography]{revtex4}

\usepackage[dvipdfm]{graphicx}
\usepackage{natbib}
\usepackage{subfigure}
\usepackage{tabularx}
\usepackage{epsfig}
\usepackage{longtable}
\usepackage{amsfonts}
\usepackage{rotating}

\usepackage{hyperref}
\usepackage{babel}

\usepackage{currfile}
\usepackage{bm}

\usepackage{color}

\newcommand{\bea}{\begin{eqnarray}}
\newcommand{\eea}{\end{eqnarray}}
\newcommand{\la}{\label}
\newcommand{\be}{\begin{equation}}
\newcommand{\ee}{\end{equation}}



\begin{document}

\title{On the effective hydrodynamics of FQHE
 }

\author{Alexander G. Abanov}
\affiliation{Department of Physics and Astronomy, Stony Brook University,
Stony Brook, NY 11794-3800}
\affiliation{Simons Center for Geometry and Physics, Stony Brook University,
Stony Brook, NY 11794-3636}

\date{\today}

\begin{abstract}

We present here a classical hydrodynamic model of a two-dimensional fluid which has many properties of Fractional Quantum Hall effect. This model incorporates the FQHE relation between the vorticity and density of the fluid and  exhibits the Hall viscosity and Hall conductivity found in FQHE liquids. We describe the relation of the model to the Chern-Simons-Ginzburg-Landau theory of FQHE and show how Laughlin's wave function is annihilated by the quantum velocity operator.


\end{abstract}

\maketitle


\paragraph{1.~Introduction.}
The goal of this paper is to give a simple classical hydrodynamic model of two-dimensionl ideal fluid which exhibits the properties usually associated with Quantum Hall states. The idea of describing these states as in terms of the effective field theory goes back to the seminal paper \cite{1986-GirvinMacDonaldPlatzman} and culminated in the effective Chern-Simons-Ginzburg-Landau theory (CSGL) of Fractional Quantum Hall effect (FQHE) \cite{1989-Read,1989-ZHK}. The CSGL description was shown to capture many properties of FQHE states (for review see \cite{1992-Zhang}). It was also noted that CSGL can be rewritten in terms of equations of hydrodynamic type describing the dynamics of FQHE liquid \cite{1990-Stone}. It was recognized in \cite{1990-Stone} that the main feature of the effective hydrodynamic formulation of FQHE fluids is the relation between density of the fluid and its vorticity which will be referred to as FQHE constraint. 

The hydrodynamics of \cite{1990-Stone} possesses many features of FQHE states. However, both the hydrodynamic formulation of \cite{1990-Stone} and CSGL descriptions fail to give the correct value for the Hall viscosity of FQHE states. In this paper we correct the hydrodynamic description of FQHE (and CSGL description) to allow for the non-vanishing Hall viscosity existing in FQHE states.

The Hall viscosity is the dissipationless part of the viscosity tensor. It is allowed in isotropic two-dimensional fluids if the parity symmetry is broken. In the context of FQHE it was introduced in \cite{1995-AvronSeilerZograf}  relatively recently \footnote{The terms of Hall viscosity type were known in the dynamics of plasma in magnetic field for some time \cite{LandauLifshitz-10}.} and then considered in many works  \cite{1998-Avron,2007-TokatlyVignale,2009-TokatlyVignale-JPhys,2009-TokatlyVignale-Erratum,2009-Read-HallViscosity,2009-Haldane-HallViscosity,2011-ReadRezayi,bradlyn2012kubo}. The term ``Hall viscosity'' was introduced in \cite{2009-Read-HallViscosity}. It is also known as ``odd viscosity'' and ``Lorentz shear modulus''.

In the following we introduce the classical action for effective hydrodynamics and study its properties. Then we compare the obtained properties with known properties of the FQHE fluid. In this paper we consider only bulk hydrodynamics of FQHE fluid and do not pay attention to (very important) boundary terms. 

\paragraph{2.~Hydrodynamic action.}
We start with the following phase space action written in terms of two scalar fields: the particle density of the 2d fluid $\rho(x,t)$ and the additional scalar potential $\vartheta(x,t)$ conjugated to the density 
\bea
	L &=&  -\rho \left[\partial_t\vartheta +eA_{0}
	+\frac{m\bm{v}_\alpha^2}{2} +(\xi-\alpha)\omega_c \right]
 \nonumber \\
 	&-& V_{\alpha}(\rho)
	-2\frac{\gamma-\alpha^2}{m} (\bm{\nabla}\sqrt{\rho})^{2} \,,
 \la{effhydroL} \\
 	V_\alpha (\rho) &=& V(\rho) +\alpha \frac{2\pi\beta}{m}\rho^2\,.
\eea
Here the auxiliary field $\bm{v}_\alpha$ is \textit{defined} as
\bea
 	m \bm{v}_\alpha &=& \bm{\nabla}\vartheta+\frac{e}{c}\bm{A}
	+\beta\bm{\nabla}^{*}\rho^L+\alpha\bm{\nabla}^{*}\ln\rho\,,
 \la{defv} 
\eea
with logarithmic integral transform
\bea
	f^L(\bm{r}) &=& 2\pi \Delta^{-1}f
	= \int d^{2}\bm{r}'\; \ln|\bm{r}-\bm{r}'|\, f(\bm{r}')\,,
 \la{defphi} 
\eea
where we assumed that $f(\bm{r})$ decays fast at spatial infinity.
The star notation in (\ref{defv}) is specific for two dimensions and means $a_i^*=\epsilon^{ij}a_j$ for any 2d vector $\bm{a}$.
In (\ref{effhydroL}) the function $V(\rho)$ is the function of density and the $\gamma$-term describes the corrections to the action quadratic in density gradients. We have introduced constants $\xi$, $\gamma$ and $\beta$ physical meaning of which will be identified later.\footnote{All these constants are measured in the same units as $\hbar$.} The parameter $\alpha$ is auxiliary. Indeed, collecting all $\alpha$ dependent terms of (\ref{effhydroL}) we obtain the complete derivative $\alpha \bm{\nabla}^*(\rho \bm{v}_{\alpha=0})$ which does not change bulk equations of motion in classical theory.  The external electromagnetic potentials $A_0$ and $\bm{A}$ are related to electric and magnetic fields as $B=\bm{\nabla}\times\bm{A}$ and $\bm{E}=-\bm{\nabla}A_0+c^{-1}\partial_t\bm{A}$ and the local cyclotron frequency is defined as $\omega_c=eB/(mc)$.

The variation of the action (\ref{effhydroL}) with respect to the vector potential $\bm{A}$ gives the expression of electromagnetic current and charge density in terms of fundamental variables of the theory
\bea
	\bm{j}_e &=& -c\frac{\partial L}{\partial \bm{A}} 
	= e\rho \bm{v}_{\xi}\,,
	\quad
	\rho_e = -\frac{\partial L}{\partial A_0} =e\rho\,.
 \la{current} 
\eea
We assume that the fluid described by the effective action (\ref{effhydroL}) is made out of one species of constituent particles (spin polarized electrons) with electric charge $e$ and mass $m$. Therefore, we identify the electric current density (\ref{current}) with the momentum density $\bm{j}_\xi$ of the fluid as
\bea
	\bm{j}_\xi=m\bm{j}_e/e=m\rho\bm{v}_{\xi}\,.
 \la{currentm}
\eea

The scalar field $\vartheta$ is given by $\delta S/\delta (\partial_t\rho)$ and should be considered as a field conjugated to the particle density $\rho$. In particular, their Poisson's brackets are given by
\bea
	\left\{\vartheta(x),\rho(x')\right\} =\delta(x-x')\,.
\eea
The Hamiltonian corresponding to (\ref{effhydroL}) can be easily found as
\bea
	H &=& \int d^{2}x\, \left[\frac{\bm{j}_\xi^2}{2m\rho} +e\rho A_0+V_\xi(\rho)
	+2\frac{\gamma-\xi^2}{m} (\bm{\nabla}\sqrt{\rho})^{2}\right]\,.
 \la{hamdef}
\eea
The variation of the Hamiltonian with respect to the momentum density gives $\delta H/\delta \bm{j}_\xi=\bm{v}_\xi$ which allows us to identify the field $\bm{v}_\xi$ conjugated to the momentum density as a proper hydrodynamic velocity of the model under consideration.

\paragraph{3.~FQHE constraint.}
We notice here that generally it is necessary to introduce Clebsch variables to formulate a variational principle for two- or three-dimensional hydrodynamics \cite{1997-ZakharovKuznetsov}. A notable exception is the hydrodynamics of a superfluid where the velocity is constrained to have a vanishing vorticity. The formulation (\ref{effhydroL}) is very close to the superfluid hydrodynamics as the velocity (\ref{defv}) satisfies the constraint
\bea
	\bm{\nabla}\times\bm{v}_\alpha &=& \omega_c 
	- \frac{2\pi\beta}{m} \rho-\frac{\alpha}{m}\Delta\ln\rho\,,
 \la{fqhe-constraint}
\eea
where we used (\ref{defphi}). The vorticity of the fluid (\ref{fqhe-constraint}) is not zero but is fully defined by the density of the fluid. The analogy of FQHE hydrodynamics with the superfluid one and the constraint (\ref{fqhe-constraint}) (without $\alpha$-term) was first noticed in \cite{1990-Stone}. We will refer to (\ref{fqhe-constraint}) as to the FQHE constraint.

Due to the FQHE constraint (\ref{fqhe-constraint}) the two-component velocity field has only one independent component which is represented by a scalar field $\vartheta$ (\ref{defv}) while the other component is fixed by the density configuration.

In deriving (\ref{fqhe-constraint}) from (\ref{defv}) we assumed that the scalar field $\vartheta$ is single-valued and $\bm{\nabla}\times\bm{\nabla}\vartheta=0$. As in the superfluid dynamics we could relax this constraint and treat $\vartheta$ as a phase field allowing for singular vortex-like configurations of the field $\bm{\nabla}\vartheta$. We will come back to these important configurations later.

\paragraph{4.~Equations of motion}
The variation of (\ref{effhydroL}) with respect to $\vartheta$ immediately gives the continuity equation for the density of the fluid
\bea
	\partial_t\rho +\bm{\nabla}(\rho\bm{v}_\alpha) &=& 0\,.
 \la{cont}
\eea
Variation with respect to the density field $\rho$ is more tricky and produces \footnote{The overall chemical potential can be absorbed into $A_0$.}
\bea
	\partial_t\vartheta &+& eA_0+\frac{\bm{v}_\alpha^2}{2} +(\xi-\alpha)\omega_c
	+\frac{\partial V_{\alpha}}{\partial \rho} 
	-2\frac{\gamma-\alpha^2}{m} \frac{\Delta\sqrt{\rho}}{\sqrt{\rho}}
 \nonumber \\
	&-& \left(\frac{\alpha}{\rho}+2\pi\beta\Delta^{-1}\right)\bm{\nabla}^*(\rho\bm{v}_\alpha) = 0 \,,
 \la{bern}
\eea
where the inverse Laplacian $\Delta^{-1}$ is defined in (\ref{defphi}).

Using (\ref{cont},\ref{bern}) and definitions (\ref{defv},\ref{defphi}) after straightforward (but tedious) transformations we obtain
\bea
	\partial_t j_{\alpha \,i} 
	&+& \partial_k\Big[m\rho v_{\alpha\, i}v_{\alpha\, k} 
	+\left(P_\alpha+(\xi-\alpha)\rho\omega_c\right)\delta_{ik} 
  \la{eqrhovalpha}\\
	&+&  \alpha\rho(\partial_i v_k^*+\partial_i^*v_k)
	-\frac{\gamma-\alpha^2}{m} \rho \partial_i\partial_k\ln\rho
	\Big]
 	= \rho F_i \,,
 \nonumber 
\eea
where  the Lorentz force \footnote{Notice that in the definition of the Lorentz force we use the velocity corresponding to the electric current (\ref{current}).}
\bea
	F_i = eE_i -\frac{e}{c}B v_{\xi\,i}^* \,.
\eea 
Here we introduced the pressure $P_\alpha(\rho)$ as  $\rho\partial_\rho^2 V_\alpha=\partial_\rho P_\alpha$. 

\paragraph{5.~Stress tensor and Hall viscosity.}
Evaluating (\ref{eqrhovalpha}) at $\alpha=\xi$ we obtain for the momentum density (\ref{currentm})
\bea
	\partial_t j_{\xi\,i} +\partial_k T_{ik} &=& \rho F_i\,,
 \la{jeq}
\eea
where the stress tensor $T_{ik}$ is defined as
\bea
	T_{ik} &=& m\rho v_{\xi\, i}v_{\xi\, k} +P_\xi\delta_{ik}  
	-\frac{\gamma-\xi^2}{m} \rho \partial_i\partial_k\ln\rho
 \nonumber \\
 	&+&  +\xi\rho(\partial_i v_{\xi\,k}^*+\partial_i^*v_{\xi\,k})\,.
 \la{stress}
\eea
The last term here is the standard form of the Hall viscosity part of the stress tensor \cite{1995-AvronSeilerZograf,1998-Avron}. It describes the additional pressure normal to the velocity of the fluid which appears in the presence of the shear. This part of the viscous stress is non-dissipative and is allowed in ideal two-dimensional fluids if parity symmetry is broken. The Hall viscosity can be read from this term as
\bea
	\eta_H &=& \xi\rho \,.
 \la{Hviscosity}
\eea
Thus we showed that the action (\ref{effhydroL}) results in the Hall viscosity (\ref{Hviscosity}). This is one of the results of this paper. Notice that the $\xi$-shift of velocity $\bm{v}_\xi$ includes an additional physics (Hall viscosity) compared to the FQHE constraint (\ref{fqhe-constraint}). It can be shown that the $\xi$ shift of velocity produces Hall viscosity in more general models of 2d hydrodynamics even in the absence of FQHE constraint.

Let us pause here to explain how the $\xi$ term in the expression for the hydrodynamic velocity $\bm{v}_\xi$ (\ref{defv})  might appear from microscopic considerations. We consider a 2d fluid made out of point particles each having an internal angular momentum equal $\mu$. Our goal is to describe this fluid by its density and velocity fields. Notice that we restrict ourselves to the description which does not include any additional degrees of freedom. Therefore, the effective hydrodynamic description we are looking for should imitate possible redistributions of angular momentum density of the fluid due to varying density of particles. The angular momentum of the finite volume of the fluid is given then by $\int d^{2}x\,(\bm{r}\times \rho\tilde{\bm{v}}+\mu \rho)$, where $\tilde{\bm{v}}$ is the microscopic velocity of particles. Integrating this expression by parts and neglecting surface terms we obtain $\int d^{2}x\,\bm{r}\times \left(\rho\tilde{\bm{v}}+\frac{\mu}{2} \bm{\nabla}^{*}\rho\right)$. We observe that the angular momentum of the fluid originating from the intrinsic momentum of constituent particles can be imitated by the $\xi$ term in  (\ref{defv}) with $\xi = \mu/2$. 
Of course, the correspondent gradient correction of current $\rho\tilde{\bm{v}}\to \rho\tilde{\bm{v}}+\frac{\mu}{2}\bm{\nabla}^*\rho$ is a well known magnetization current. Thus, in this simple case the origin of the Hall viscosity in (\ref{Hviscosity}) is the intrinsic angular momentum of fluid particles. The relation between the Hall viscosity and angular momentum per particle was introduced by Read \cite{2009-Read-HallViscosity}. 

\paragraph{6.~Linearization and dispersion of magnetoplasmons.}
Let us consider solutions of (\ref{effhydroL}) in linear approximation. In the absence of an external potential $A_0=0$ and in constant magnetic field $B=\bar{B}=const$ we have the following static background solution of hydrodynamics
\bea
	\rho &=& \bar\rho = \frac{m\bar\omega_c}{2\pi\beta}\,, 
	\qquad
 	\bm{v}_\xi = 0\,.
 \la{background}
\eea
Here we use bar to denote all background quantities. 

A very important feature of the background solution (\ref{background}) is that the background density is proportional to the magnetic field. In the context of FQHE we should identify $\bar\rho$ as $\nu/(2\pi l^2)$, where $\nu$ is a dimensionless filling factor and $l^2=\hbar c/(e\bar{B})$ is a magnetic length. Then
the parameter $\beta$ is proportional to the inverse filling factor of FQHE
\bea
	\beta = \hbar \nu^{-1}\,.
 \la{beta}
\eea

Let us expand the effective action (\ref{effhydroL}) in deviations from the background (\ref{background}). We introduce deviations $\delta \rho =\rho -\bar\rho$ etc and keeping only quadratic terms we obtain from (\ref{effhydroL}) in Fourier components
\bea
	L &=& \frac{1}{2} (\delta\theta^\dagger,\delta\rho^\dagger)\hat{\Gamma}
	\left(
	\begin{array}{c}
		\delta\theta \\ \delta\rho
	\end{array}
	\right)
	-\frac{m\bar{\rho}}{2k^{2}}|\delta\omega_c|^{2}
 \la{L2mod} \\
 	&+& \frac{\delta\rho^\dagger}{2k^2} 
	\left[
	-e i\bm{k}\delta\bm{E} 
	+m\bar\omega_c\delta\omega_c\left(1-\frac{\xi k^2}{m\bar\omega_c}\right)
	\right]
	+c.c.
 \nonumber 
\eea
Here $\dagger$ denotes complex conjugate and we introduced a new notation for the gauge-invariant combination
\bea
 	\delta\theta &=& \delta\vartheta-\frac{e}{c}\frac{i\bm{k}\delta\bm{A}}{k^2} 
\eea
and  
\bea
	\hat{\Gamma} &=& 
	\left(
	\begin{array}{cc}
	-\frac{\bar\rho}{m} k^2 & -i\omega \\
	i\omega & -\frac{m\Omega_{k}^{2}}{\bar{\rho}k^2}
	\end{array}
	\right) \,,
 \la{Gamma}
\eea
where
\bea
	\frac{\Omega_{k}^{2}}{\bar{\omega}_{c}^{2}} &=&
	1+\frac{mV_{\rho\rho}}{2\pi\beta} \frac{k^{2}}{m\bar{\omega}_{c}}
	+\gamma \left(\frac{k^{2}}{m\bar{\omega}_{c}}\right)^{2} \,.
 \la{Omegak}
\eea
$V_{\rho\rho}$ denotes the second derivative of $V(\rho)$ calculated at $\rho=\bar\rho$.
The condition $\det(\hat{\Gamma})=0$ gives the dispersion of small deviations from the background solution as $\omega = \pm \Omega_k$, i.e., the dispersion of magnetoplasmons. It is gapful with the gap given by the cyclotron frequency $\omega_c$ in accordance to Kohn's theorem.
The effective compressibility of the fluid is infinite because of this gap. The origin of this incompressibility is easy to understand. Indeed, because of the constraint (\ref{fqhe-constraint}) the attempt to change the density of the fluid results in the vorticity and the kinetic energy proportional to $v^{2}$ diverges \cite{1990-Stone}.

\paragraph{7.~Linear response and Hall conductivity.}
Using the definition of electric current (\ref{current}) and the linearized theory (\ref{L2mod}) it is 
straightforward to obtain linear response formulas relating the current and the density of the fluid to external weak perturbations of electric and magnetic fields. In particular, we obtain for the dynamical Hall conductivity 
\bea
	\sigma_H(\omega,k) = \frac{e^2}{2\pi\beta}\frac{\bar\omega_{c}^{2}}{\omega^2-\Omega_k^2}
	\left(1-\frac{\xi k^2}{m\bar\omega_{c}}\right) \,.
 \la{Hsigma2}
\eea
The expression (\ref{Hsigma2}) has poles at the frequencies of propagating modes (\ref{Omegak}). Expanding (\ref{Hsigma2}) in $k$ at $\omega=0$ we obtain for the static Hall conductivity 
\bea
	\sigma_H &\approx& -\frac{e^2}{2\pi\beta}
	\left[1
	+\left(\xi-\frac{mV_{\xi\,\rho\rho}}{2\pi\beta}\right)
	\frac{k^2}{m\bar\omega_c}
	+\lambda
	\left(\frac{k^2}{m\bar\omega_c}\right)^2\right]\,.
 \la{Hsigma}
\eea
where $\lambda = \xi\frac{mV_{\rho\rho}}{2\pi\beta}+\left(\frac{mV_{\rho\rho}}{2\pi\beta}\right)^{2}
	-\gamma$. We notice from (\ref{Hsigma}) that the $k^2$ correction to Hall conductivity has two terms: one related to the Hall viscosity (\ref{Hviscosity}) and the other from the intrinsic compressibility of the fluid $\kappa^{-1}=\bar\rho^2 V_{\xi\,\rho\rho}$. This is consistent to the results of \cite{hoyos2012hall}. The next $k^4$ correction is expected to be non-universal as it explicitly depends on the $\gamma$-term in (\ref{effhydroL}).

We can also find the change of density under small variations of electric and magnetic fields as
\bea
	\frac{\delta\rho}{\bar\rho} =\frac{\bar{\omega}_{c}^{2}}{\omega^2-\Omega_k^2}\left[
	\frac{e}{m\bar{\omega}_{c}^{2}}(\bm{\nabla}\bm{E}) 
	- \left(1-\frac{\xi k^2}{m\bar\omega_c}\right)\frac{\delta \omega_{c}}{\bar\omega_{c}}\right] \,.
 \la{deltarho}
\eea

So far our consideration was purely classical. Assuming that (\ref{L2mod}) can also be used in the path integral to compute quantum density fluctuations we immediately find the dynamical correlation function $\langle \delta\rho\delta\rho\rangle_{\omega,k} =-i\hbar (\hat{\Gamma}^{-1})_{22}$ corresponding to the static structure factor (same time correlation function)
\bea
	s(k) = \frac{1}{\bar\rho}\langle \delta\rho \delta\rho\rangle_{k} 
	\approx \frac{(kl)^2}{2}
	\left[1-\frac{(kl)^2}{2} \frac{mV_{\rho\rho}}{2\pi\hbar\beta} \right]\,.
 \la{sk}
\eea
where we used $\frac{\hbar k^{2}}{m\bar\omega_{c}}=(kl)^2$.

\paragraph{8.~Comparison to FQHE.}
So far we considered the hydrodynamic model (\ref{effhydroL}) as an effective theory treating all parameters phenomenologically. These parameters should be calculated from the microscopic theory. In the following we are going to consider FQHE effect as an underlying theory and will match the phenomenological parameters of (\ref{effhydroL}) against the results known for FQHE. In the following we restrict ourselves to the simplest FQHE fractions of Laughlin's type with the filling factor $\nu=1/(2n+1)$ with integer $n$.

Comparing the dc Hall conductivity at $k=0$ given by (\ref{Hsigma}) with $\sigma_H=\frac{e^2}{2\pi\hbar}\nu$ we immediately obtain the parameter $\beta$ in terms of the FQHE filling factor (\ref{beta}).

The value of Hall viscosity (\ref{Hviscosity}) should be compared to $\eta_H= \frac{\nu^{-1}}{4}\hbar\rho$ \cite{2009-Read-HallViscosity}. We have
\bea
	\xi = \hbar\nu^{-1}/4\,.
 \la{xi}
\eea
The Hall conductivity (\ref{Hsigma}) has a $k^2$ correction which can be matched to the one for FQHE \cite{hoyos2012hall} to give
\bea
	\frac{mV_{\xi\,\rho\rho}}{2\pi\beta} =\hbar
 \la{mVrhorho}
\eea
and correspondingly $\frac{mV_{\rho\rho}}{2\pi\hbar\beta}=1-2\xi/\hbar = 1-\nu^{-1}/2$. This expression being substituted into (\ref{sk}) coincides with the structure factor obtained in \cite{1986-GirvinMacDonaldPlatzman}.
We notice here that the parameters of the effective hydrodynamics (\ref{effhydroL}) are determined by the quantum physics of FQHE and have explicit dependence on $\hbar$ (\ref{beta},\ref{xi},\ref{mVrhorho}).

We can summarize the obtained linear response results for FQHE as a ``minimal'' effective hydrodynamics for FQHE given by the Hamiltonian
\bea
	H &=& \int d^{2}x\,\rho \left[\frac{m\tilde{\bm{v}}^2}{2} +\frac{\hbar\omega_{c}}{2}+eA_0 \right]\,,
 \la{hamFQHE} \\
 	\frac{m\tilde{\bm{v}} }{\hbar}&=& \bm{\nabla}\frac{\vartheta}{\hbar}+\frac{e\bm{A}}{\hbar c}
	+\nu^{-1}\bm{\nabla}^{*}\rho^L+\frac{\nu^{-1}-2}{4}\bm{\nabla}^{*}\ln\rho\,.
 \la{tildev}
\eea
This choice of the form of the Hamiltonian corresponds to the choice of the parameter $\alpha=\hbar \frac{\nu^{-1}-2}{4}$.
The second term of (\ref{hamFQHE}) does not affect equations of motion in constant magnetic field. However, it is very important to keep it as it gives a way to correctly identify the current (\ref{current}) and, therefore, the momentum density in the case of electronic fluid. This term can also be interpreted as the contribution of zero mode motion of the quantum theory to the effective classical formulation (\ref{hamFQHE}).

It is important that although we have used the mass of the electron to formulate the theory, the chosen values of parameters reproduce linear responses
\bea
	\eta_{H} &=& \frac{\nu^{-1}}{4}\hbar\bar\rho\,,
 \\
 	\sigma_{H} &\approx & - \frac{e^{2}}{2\pi\hbar}\nu 
	\left[1+\frac{\nu^{-1}-4}{4}(kl)^{2}\right]\,,
 \\
 	s(k) &\approx &
	\frac{(kl)^2}{2}
	\left[1+\frac{\nu^{-1}-2}{4}(kl)^2 \right]
\eea
that do not depend on $m$ up to the given orders in $k$.

\paragraph{9.~Quasiholes and energy gap.}
So far we worked only with linear response properties of the effective hydrodynamics. However, the hydrodynamic model (\ref{effhydroL}) is nonlinear and it is very interesting to study its essentially nonlinear consequences. Let us take a look at vortex-like solutions of the model. 

We consider the constant background magnetic field and relax the requirement of single-valuedness of $\vartheta$. Taking $\vartheta =-C\arg(\bm{r})$ and choosing the asympotics at $\bm{r}\to\infty$ as $\delta\rho^L \to -\frac{C}{\beta}\ln(r)$ we find $\tilde{\bm{v}}\to 0$ and obtain the solution with finite energy (\ref{hamFQHE}). The exact form of this solution at small distances depends on the shape of the potential $V(\rho)$ and is determined by the microscopic physics.  The obtained energy of the solution determines the gap separating the ground state of FQHE from the lowest quasihole states. \footnote{In simplest models of FQHE one expects this energy of the order of $e^{2}/l$ as determined by the 3d Coulomb scale.} It defines also the phenomenological parameter $m^{*}$ \cite{1989-Read,1992-Zhang}

For the total local charge of the solution we have $\int dx\, e\delta\rho =-\nu e C/\hbar$. The latter relation is obtained by taking the value (\ref{beta}) for FQHE fluid. We identify this solution as Laughlin's quasihole and require that $C$ is a multiple of $\hbar$ to fix the correct quasihole elementary charge of $-e\nu$. The field $\vartheta/\hbar$ is multivalued and changes by multiples of $2\pi$ around the core of the quasihole. This is possible if $\vartheta/\hbar$ is a compact field understood as a phase of a single-valued order parameter of FQHE.

\paragraph{10.~Chern-Simons-Ginzburg-Landau theory and effective hydrodynamics.} It is easy to show by straightforward calculation that the model (\ref{effhydroL}) can be obtained from the following modified Chern-Simons-Ginzburg-Landau theory (CSGL)
\bea
	L &=& \Phi^*\left\{i\hbar\partial_t-eA_0-a_0 
	-\frac{\xi}{m} \bm\nabla\times\left(\bm{a}+\frac{e}{c}\bm{A}\right) \right\}\Phi
 \nonumber \\
 	&-& \frac{1}{2m}\left|\left(-i\hbar\bm{\nabla}+\bm{a}+\frac{e}{c}\bm{A}\right)\Phi\right|^2
	-\frac{1}{4\pi\beta}\epsilon^{\mu\nu\lambda}a_\mu \partial_\nu a_\lambda
 \nonumber \\
 	&-& V_\xi(\rho) -\frac{2\gamma-\hbar^2/2}{m}(\bm{\nabla}\sqrt{\rho})^2\,,
 \la{CSGL}
\eea
where $\Phi$ is a complex scalar field and $\rho=|\Phi|^2$. Indeed, ``integrating out'' the statistical gauge field $a_0,\,\bm{a}$ and making a hydrodynamic substitution $\Phi=\sqrt{\rho}e^{i \vartheta/\hbar}$ in (\ref{CSGL}) we arrive to (\ref{effhydroL}). In particular, the variation of (\ref{CSGL}) over $a_0$ gives $\bm{a}=\beta\bm{\nabla}^*\rho^{L}$. 

The model (\ref{CSGL}) is slightly modified compared to \cite{1989-Read,1989-ZHK,1992-Zhang}. The main difference is the last term in the first line. This term proportional to the parameter $\xi$ corrects the original CSGL so that the equivalent hydrodynamic model (\ref{effhydroL}) exhibits the Hall viscosity (\ref{Hviscosity}). The hydrodynamic formulation of (\ref{CSGL}) was studied in \cite{1990-Stone} but without $\xi$ term. The addition of this term to (CSGL) is the main result of this paper.

\paragraph{11.~Quantum velocity operator and Laughlin's wave function.} Matching phenomenological parameters of (\ref{effhydroL}) with calculations made for Laughlin's state of FQHE we obtained that the minimal classical model describing some FQHE phenomenology  (\ref{hamFQHE},\ref{tildev}). Here we consider the expression (\ref{tildev}) for the modified velocity and argue that its quantized version annihilates Laughlin's function in density representation. The quantized (naively) version of (\ref{tildev}) is given by
\bea
	\tilde{v} = 2i\frac{\hbar}{m}\partial\left[- \frac{\delta}{\delta \rho} +\frac{|z|^2}{4l^2}
	+\nu^{-1}\rho^{L} +\frac{\nu^{-1}-2}{4}\ln\rho \right]\,,
 \la{qv}
\eea
where we introduced complex notations $z=x+iy$, $2\partial=\partial_x-i\partial_y$, $\tilde{v}=\tilde{v}_x-i\tilde{v}_y$ and used the radial gauge.
On the other hand the Laughlin's wave function for $\nu=1/n$ after singular gauge transformation $\Psi_L = \prod_{i<j} |z_i-z_j|^n e^{-\sum_i \frac{|z_i|^2}{4l^2}}$ can be rewritten as a functional of density $\rho$ 
\bea
	\Psi_L[\rho] &=& \exp\left\{\frac{\nu^{-1}}{2}\int d^2z\,d^2z'\, \rho\ln|z-z'| \rho'\right.
\la{psirho} \\
	&-& \left. \int d^2z\, \rho\frac{|z|^2}{4l^2}
	+\frac{\nu^{-1}-2}{4}\int d^2z\,\rho\ln \rho \right\}\,,
 \nonumber
\eea
where $\rho'=\rho(z')$ etc. It is easy to see that the quantized velocity operator (\ref{qv}) annihilates (\ref{psirho}). The main content of this work is the inclusion of the last term of (\ref{psirho}) known in FQHE \cite{2006-ZabrodinWiegmann} into the classical hydrodynamics (\ref{hamFQHE},\ref{tildev}).

\paragraph{12.~Conclusion.} To conclude, we formulated the effective classical hydrodynamics (\ref{hamFQHE}) which possesses many properties associated with collective dynamics of FQHE fluid. We did not seriously consider the quantum hydrodynamic version of the classical model but showed that the naively quantized velocity operator (\ref{qv}) annihilates the Laughlin's wave function written in collective form. To describe fully the physics of FQHE the hydrodynamic model we considered should be quantized and projected to the first Landau level. As a result of this projection the modes corresponding to (\ref{Omegak}) will be frozen and only intra-Landau level dynamics will remain. We are planning to address these issues elsewhere. Finally, it is hard to overlook the resemblance of all expressions of this work to the corresponding formulas for the collective description of the Calogero-Sutherland model \cite{2005-AbanovWiegmann,2006-BettelheimAbanovWiegmann-shocks,2009-AbanovBettelheimWiegmann}.


\paragraph{Acknowledgment.}
This study had started as a common project with P. Wiegmann. It turned out that our methods and results were complimentary and we decided to present them in separate publications, see \cite{wiegmann2012quantum}. I am grateful to N. Read for discussions of Hall viscosity.
The work was supported by the NSF under Grant No.\ DMR-1206790.


\bibliographystyle{my-refs}



\bibliography{abanov-bibliography}

%


\end{document}